\documentclass[12pt]{article}
\usepackage{epsf}
\usepackage{amsmath}
\usepackage{graphics}
\usepackage{cite}

\renewcommand{\thefootnote}{\#\arabic{footnote}}
\begin{document}

\newcommand{\gtrsim}{ \mathop{}_{\textstyle \sim}^{\textstyle >} }
\newcommand{\lesssim}{ \mathop{}_{\textstyle \sim}^{\textstyle <} }

\newcommand{\rem}[1]{{\bf #1}}

\renewcommand{\thefootnote}{\fnsymbol{footnote}}
\setcounter{footnote}{0}
\begin{titlepage}

\def\thefootnote{\fnsymbol{footnote}}

\begin{center}

\vskip .5in
\bigskip
\bigskip
{\Large \bf A Note on Embedding Nonabelian Finite Flavor Groups in Continuous
Groups}

\vskip .45in

{\bf Paul H. Frampton $ ^{(a)} $ \footnote{frampton@physics.unc.edu}
Thomas W. Kephart$^{(b)}$\footnote{tom.kephart@gmail.com}
Ryan M. Rohm$^{(a)}$\footnote{rmrohm@physics.unc.edu}}

\bigskip
\bigskip

{$^{(a)}$ Department of Physics and Astronomy, UNC-Chapel Hill, NC 27599.}

{$^{(b)}$ Department of Physics and Astronomy, Vanderbilt University,
Nashville, TN 37235.}

\end{center}

\vskip .4in

\begin{abstract}
A nonabelian finite flavor group $G \subset SO(3)$ can have 
double covering $G^{'} \subset SU(2)$ such that $G \not\subset G^{'}$.
This situation is not contradictory, but quite natural, and we give
explicit examples such as $G=D_n, G^{'}=Q_{2n}$ and
$G=T, G^{'}=T^{'}$. This observation can be crucial in particle theory model building.
\end{abstract}

\end{titlepage}

\renewcommand{\thepage}{\arabic{page}}
\setcounter{page}{1}
\renewcommand{\thefootnote}{\#\arabic{footnote}}

\newpage

\noindent {\it Introduction.}

\bigskip
\bigskip

\noindent One promising direction for extending the standard model of particle
 theory
lies in using a flavor symmetry  $G_F$ which commutes with the gauge group. 
The flavor symmetry can lead to predictions for the quarks and lepton mixing angles.
Typically $G_F$ is a 
finite nonabelian group.

\bigskip

\noindent The standard model gauge group $G_{SM}$ is comprised of the
direct product of continuous groups $G_{SM} \equiv SU(3) \times SU(2) \times U(1)$.
The flavor group $G_F$
is itself often a subgroup of a Lie group: $G_F \subset G_L$ 
Here we focus on the cases $G_L=SO(3)$ and $G_L=SU(2)$.

\bigskip

\noindent There can be, as we shall discuss, a pair of candidates 
$G^{i}_F$ ($i = 1, 2$) which have the following
set of seemingly contradictory properties.

\bigskip

\begin{itemize}

\item $G_F^{1} \subset SO(3)$

\item $G_F^{2} \subset SU(2)$

\item $G_F^{2}$ is the double covering of $G_F^{1}$
just as $SU(2)$  for $SO(3)$.

\item $G_F^{1} \not\subset G_F^{2}$

\end{itemize}

\bigskip

\noindent The apparent contradiction in the last statement is only
that, since $SO(3) \not\subset SU(2)$! The situation described is
not uncommon.  Indeed, we will show that $G_F^{1} \subset G_F^{2}$
only when $G_F^{1} $ is abelian.  First we will set the stage by
considering a few explicit examples, including an infinite series of
$SU(2)$ subgroups, after a brief and self-contained survey of the
small nonabelian groups.

\bigskip

\noindent {\it  Nonabelian groups with order $g \le 31$}

\bigskip

\noindent Some of our examples will be taken from nonabelian groups
with order $g \leq 31$.  There are 45 such groups.  Of the nonabelian
finite groups, the best known are perhaps the permutation groups $S_N$
(with $N \geq 3$) of order $N!$ The smallest non-abelian group is
$S_3$ ($\equiv D_3$), the symmetry of an equilateral triangle with
respect to all rotations in a three dimensional sense. This group is a
member of two disparate infinite series, $S_N$ and $D_N$, distinct
for $N > 3$.  Both have elementary geometrical significance since
the symmetric permutation group $S_N$ is the symmetry of the N-plex in
N dimensions while the dihedral group $D_N$ is the symmetry including
inversion of the planar N-agon in 3 dimensions.

\bigskip
\bigskip

\noindent As a family symmetry, the $S_N$ series (and the even-parity
series $A_N$) becomes less interesting as the order and the
dimensions of the representions increase, because $S_N$ contains
transformations which independently exchange elements of the
fundamental $N$-plet.  Only $S_3$ and $S_4$ are of any interest as
symmetries associated with the particle spectrum\cite{Pak}. Also, the
order (number of elements) of the $S_N$ groups grow factorially with
N, and the order of the dihedral groups increase only linearly with N and
their irreducible representations are all one- and two-
dimensional. This is reminiscent of the representations of the
electroweak $SU(2)_L$ used in Nature.

\bigskip

\noindent Each $D_N$ is a subgroup of $O(3)$ and has a counterpart
double dihedral group $Q_{2N}$, of order $4N$, which is a subgroup of
the double covering $SU(2)$ of $SO(3)$.

\bigskip

\noindent With only the use of $D_N$, $Q_{2N}$, $S_N$ and the
tetrahedral group $T = A_4$ ( of order 12, the even permutations
subgroup of $S_4$ ), alone and in direct products with abelian groups,
we find 32 of the 45 nonabelian groups up to order 31.  (Note that
$D_6 \simeq Z_2 \times D_3, D_{10} \simeq Z_2 \times D_5$ and $ D_{14}
\simeq Z_2 \times D_7$ )

\bigskip

$$\begin{tabular}{||c||c||} \hline
g    &           \\          \hline
$6$  &  $D_3 \equiv S_3$    \\  \hline
$8$  &    $ D_4 , Q = Q_4 $  \\   \hline
$10$  &       $D_5$           \\   \hline
$12$  &  $D_6, Q_6, T$         \\   \hline
$14$  &     $D_7$               \\   \hline
$16$ &    $D_8, Q_8, Z_2 \times D_4, Z_2 \times Q$\\  \hline
$18$ &                $ D_9, Z_3 \times D_3$       \\  \hline
$20$ &       $  D_{10}, Q_{10}$ \\      \hline
$22$ &              $   D_{11}$  \\      \hline
$24$ & $D_{12}, Q_{12}, Z_2 \times D_6, Z_2 \times Q_6, Z_2 \times T$,\\  \hline
     &    $Z_3 \times D_4, D_3 \times Q, Z_4 \times D_3, S_4$\\      \hline
$26$ &     $D_{13}$                 \\  \hline
 $28 $ &     $D_{14}, Q_{14}$        \\  \hline
$30  $             & $D_{15}, D_5 \times Z_3, D_3 \times Z_5$\\  \hline
 \end{tabular}$$

\bigskip

\noindent There remain thirteen others formed by ``twisted products''
of abelian factors.  (Twisted products are a way of constructing certain
group extensions; a group extension is a way of describing a group in 
terms of a normal subgroup and its factor group.  We will see later that
double covers are also a form of group extension.)  Only certain
such ``twistings'' are permissable, namely (completing all $g \leq 31$ )

\bigskip

 $$\begin{tabular}{||c||c||}   \hline
 g & \\    \hline
 $16$  & $Z_2 \tilde{\times} Z_8$ (two, excluding $D_8$), $Z_4 \tilde{\times}
 Z_4, Z_2 \tilde{\times}(Z_2 \times Z_4)$
 (two)\\  \hline
 $18$ & $Z_2 \tilde{\times} (Z_3 \times Z_3)$\\    \hline
 $20$&  $Z_4 \tilde{\times} Z_7$ \\   \hline
 $21$&  $Z_3 \tilde{\times} Z_7$ \\    \hline
 $24$&  $Z_3 \tilde{\times} Q, Z_3 \tilde{\times} Z_8, Z_3 \tilde{\times} D_4$
 \\  \hline
 $27$&  $ Z_9 \tilde{\times} Z_3, Z_3 \tilde{\times} (Z_3 \times Z_3)$ \\
 \hline
 \end{tabular}$$

\bigskip

\noindent It can be shown that these thirteen exhaust the
classification of {\it all} inequivalent nonabelian groups up to order
thirty-one\cite{books}.

\bigskip

\noindent Of these 45 nonabelian groups, the dihedrals ($D_N$) and
double dihedrals ($Q_{2N}$), of order 2N and 4N respectively, form the
simplest sequences. In particular, they can be realized as subgroups
of $O(3)$ and $SU(2)$ respectively, the two simplest nonabelian
continuous groups.  A finite group is completely described by its
presentation as a multiplication table, but it is often more convenient to use a more
compact description.  One equivalent and compact representation is the
character table, and for physical applications this is preferred
because it also summarizes the vector representations of the group.
For the $D_N$ and $Q_{2N}$, the tables for Kronecker
products, as derivable
from the character tables, are simple to express in
general\cite{FramptonKephart}.  These tables, whose role is to 
describe the decomposition of
tensor products of irreducible vector representations (irreps) as a
direct sum of irreps, may be degenerate among different finite groups
which have distinct character tables.  The character table,
on the other hand,
just like the presentation as a multiplication table
contains the complete information about a group.

\bigskip

\noindent {\it SU(2) as the double cover of SO(3)}

\bigskip

\noindent 
Continuous groups which have the same structure for small
transformations, described by the Lie algebra, can have various global
forms distinguished by their discrete subgroups.  For compact Lie
groups, each simple Lie algebra corresponds to one global form which
is connected and simply-connected; other global forms are obtained 
through taking a quotient by some subgroup of the center of the group,
constructing an extension by a discrete symmetry of the group which
is not continuously related to the identity, or some combination of
such operations.

In particular, since $SU(2)$ is the double cover of $SO(3)$, their Lie
algebras are isomorphic.  Conversely, $SO(3)$ is obtained from $SU(2)$
if we identify pairs of elements which differ by a $2\pi$ rotation.
The irreps of $SO(3)$ have dimensions equal to odd integers $1, 3, 5,
...$ while the irreps of $SU(2)$ appear with all postive integers $1,
2, 3, ...$. In particular $SU(2)$ has defining irrep the {\bf 2} which
is a spinor and the corresponding rotations change sign after any odd
number of $2\pi$ rotations becoming the identity only after an integer
multiple of $4\pi$ rotations. This is what we mean by saying that
$SU(2)$ is the double cover of $SO(3)$: the group manifold is
topologically $S^3/Z_2$ while that of $SU(2)$ is $S^3$.

\bigskip

\noindent {\it Finite subgroups of $SO(3)$ and $SU(2)$}

\bigskip

\noindent Let us consider an explicit example. The
tetrahedral group $T$ is the symmetry of a regular
tetrahedron. As such it is a subgroup of the rotations
in three space dimensions: $T \subset SO(3)$.
From the visualization it is isomorphic to the
even permutations of four objects so $T \subset S_4$.
The irreps of $T$ are $1_1, 1_2, 1_3, 3$.

\bigskip

\noindent Next consider the binary tetrahedral
group $T^{'}$. This group is more challenging
to visualize as a symmetry. From standard treatises
we learn that $T^{'}$ is the double cover of $T$.
The irreps of $T^{'}$ are $1_1, 1_2, 1_3, 2_1, 2_2, 2_3, 3$
with the table for Kronecker products shown in Table 1.

\bigskip

\begin{center}

{\bf Table 1}

\bigskip

Table for Kronecker products of $T^{'}$ Irreps. 

\bigskip
\bigskip

\begin{tabular}{|c||c|c|c|c|c|c|c|} 
\hline
$\otimes $ & 1$_{1}$ & 1$_{2}$ & 1$_{3}$ & 2$_{1}$ & 2$_{2}$ & 2$_{3}$ &
3
\\ \hline\hline
1$_{1}$ & 1$_{1}$ & 1$_{2}$ & 1$_{3}$ & 2$_{1}$ & 2$_{2}$ & 2$_{3}$ & 3
\\
\hline
1$_{2}$ & 1$_{2}$ & 1$_{3}$ & 1$_{1}$ & 2$_{2}$ & 2$_{3}$ & 2$_{1}$ & 3
\\
\hline
1$_{3}$ & 1$_{3}$ & 1$_{1}$ & 1$_{2}$ & 2$_{3}$ & 2$_{1}$ & 2$_{2}$ & 3
\\
\hline
2$_{1}$ & 2$_{1}$ & 2$_{2}$ & 2$_{3}$ & $1+3$ & $1^{\prime }+3$ &
$1^{\prime
\prime }+3$ & $2_{1}+2_{2}+2_{3}$ \\ \hline
2$_{2}$ & 2$_{2}$ & 2$_{3}$ & 2$_{1}$ & $1^{\prime }+3$ & $1^{\prime
\prime
}+3$ & $1+3$ & $2_{1}+2_{2}+2_{3}$ \\ \hline
2$_{3}$ & 2$_{3}$ & 2$_{1}$ & 2$_{2}$ & $1^{\prime \prime }+3$ & $1+3$ &
$%
1^{\prime }+3$ & $2_{1}+2_{2}+2_{3}$ \\ \hline
3 & 3 & 3 & 3 & $2_{1}+2_{2}+2_{3}$ & $2_{1}+2_{2}+2_{3}$ & $%
2_{1}+2_{2}+2_{3}$ & 1$_{1}+1_{2}+1_{3\ }+3+3$ \\ \hline
\end{tabular}

\end{center}

\bigskip

\bigskip

\noindent We note that the group $T$ has irreps $1_1, 1_2, 1_3, 3$
which are a subset of those of $T^{'}$ and that their table for Kronecker
products shown in Table 2 is contained within that (Table 1) for $T^{'}$.

\bigskip

\begin{center}

{\bf Table 2}

\bigskip

Table for Kronecker products of $T$ Irreps.

\bigskip

\bigskip

\begin{tabular}{|c||c|c|c|c|}
\hline
$\otimes $ & 1 & $1^{\prime }$ & $1^{\prime \prime }$ & 3 \\
\hline\hline
1 & 1 & $1^{\prime }$ & $1^{\prime \prime }$ & 3 \\ \hline
$1^{\prime }$ & $1^{\prime }$ & $1^{\prime \prime }$ & 1 & 3 \\ \hline 
$1^{\prime \prime }$ & $1^{\prime \prime }$ & 1 & $1^{\prime }$ & 3 \\
\hline
3 & 3 & 3 & 3 & $1+1^{\prime }+1^{\prime \prime }+3+3$ \\ \hline
\end{tabular}

\end{center}

\bigskip

\bigskip

\noindent With all this circumstantial evidence the reader
might
well suspect from a physical perspective that $T \subset T^{'}$.
The present paper will show, however,
that this is interestingly
incorrect and will discuss in detail why $T \not\subset T^{'}$.
This can first be straightforwardly confirmed by using the representation
of T in SO(3) to construct the corresponding rotations in SU(2).
In SU(2) these elements do not close by themselves, because a $2\pi$
rotation is $-1$.  Instead, $T^{'}$ has two subsets of elements which
correspond to $T$ and $-T$ (pointwise).
\bigskip

\noindent From a more mathematical point of view, the  binary tetrahedral 
group $T^{'}$,  and the tetrahedral group $T$ live in a short exact sequence
$$1 \rightarrow Z_2 \rightarrow T\, ' \rightarrow T \rightarrow 1$$
which does not split, hence $T\, '$ is not a semidirect product 
of $Z_2 \times T$, i.e., $T^{'}$ has no subgroup isomorphic to $T$ \cite{RC}.

\bigskip

\noindent Upon further scrutiny we realize that this is also
a reflection of the following property. 
Since the group $T$ has no 2-dimensional irrep,
its embedding in $SU(2)$ can only be such that the 
defining {\bf 2} of $SU(2)$ contains two one-dimensional
representations of $T$. In that case, 
only an abelian subgroup of $T$ is represented (because any
sum of one-dimensional representations is abelian),  and all
irreps of $SU(2)$ will also contain only one-dimensional
irreps of $T$. There can exist no nontrivial embedding
of $T$ in $SU(2)$. Instead $T$ fits perfectly in $SO(3)$
as expected. Similarly $T^{'}$ cannot be embedded in $SO(3)$
but embeds regularly in $SU(2)$.

\bigskip

\noindent For the purposes of particle theory model building
$T^{'}$ {\it acts as if} it contains $T$ as a subgroup only
because Table 1 contains Table 2. Strictly speaking 
$T \not\subset T^{'}$ and any group properties which
are more subtle and which must rely on characters
will possibly reflect this crucial fact.
From another perspective
$T^{'}$ 
{\it acts as if} it contains $T$ as a subgroup only
because $T^{'}$ is a central extension of $T$. The element $-1$ is
added to $T$  in a nontrivial and consistent way, such that it 
commutes with all of the elements of $T$, and the order of each
element is doubled.

\bigskip

\noindent Is this a peculiar property only
of $SO(3) \supset T \not\subset T^{'} \subset SU(2)$?
One remark may help clarify the possibilities: 
$SU(2)$ has exactly one element of order 2 (which is the matrix 
$-1$).  Therefore it cannot contain  any subgroup with more than one
element of order 2 (which rules out all of the dihedral groups), or 
any subgroup whose element of order 2 does not commute with other elements.

\bigskip

\noindent We may quickly cite an infinite number of further
examples. The dihedral groups $D_n$ are subgroups of $SO(3)$
and their double covers are the dicyclic groups $Q_{2n}$
which are subgroups of $SU(2)$. The irreps
for {\it e.g.} $Q_8$ are $1_1, 1_2, 1_3, 1_4, 2_1, 2_2, 2_3$
with the table for Kronecker products given in Table 3.

\bigskip
\bigskip

\newpage

\begin{center}

{\bf Table 3}

\bigskip

Table for Kronecker products of $Q_8$ Irreps.

\bigskip

\bigskip

\begin{tabular}{|c||c|c|c|c|c|c|c|}
\hline
$\otimes $ & 1$_{1}$ & 1$_{2}$ & 1$_{3}$ & 1$_{4}$ & 2$_{1}$ & 2$_{2}$ &
2$%
_{3}$ \\ \hline\hline
1$_{1}$ & 1$_{1}$ & 1$_{2}$ & 1$_{3}$ & 1$_{4}$ & 2$_{1}$ & 2$_{2}$ &
2$_{3}$
\\ \hline
1$_{2}$ & 1$_{2}$ & 1$_{1}$ & 1$_{4}$ & 1$_{3}$ & 2$_{3}$ & 2$_{2}$ &
2$_{1}$
\\ \hline
1$_{3}$ & 1$_{3}$ & 1$_{4}$ & 1$_{1}$ & 1$_{2}$ & 2$_{1}$ & 2$_{2}$ &
2$_{3}$
\\ \hline
1$_{4}$ & 1$_{4}$ & 1$_{3}$ & 1$_{2}$ & 1$_{1}$ & $2_{3}$ & 2$_{2}$ &
2$_{1}$
\\ \hline
2$_{1}$ & 2$_{1}$ & 2$_{3}$ & 2$_{1}$ & 2$_{3}$ & $1_{1}+1_{3\ }+2_{2}$
& $%
2_{1}+2_{3}$ & $1_{2}+1_{4}+2_{2}$ \\ \hline
2$_{2}$ & 2$_{2}$ & 2$_{2}$ & 2$_{2}$ & 2$_{2}$ & $2_{1}+2_{3}$ & $%
1_{1}+1_{2}+1_{3}+1_{4}$ & $2_{1}+2_{3}$ \\ \hline
2$_{3}$ & 2$_{3}$ & 2$_{1}$ & 2$_{3}$ & 2$_{1}$ & $1_{2}+1_{4}+2_{2}$ &
$%
2_{1}+2_{3}$ & $1_{1}+1_{3\ }+2_{2}$ \\ \hline
\end{tabular}

\end{center}

\bigskip
\bigskip

\noindent For comparison we note that $Q_8$ is the double cover
of $D_4$ which has irreps $1_1, 1_2, 1_3, 1_4, 2$. Their
table for Kronecker products is shown as Table 4.

\bigskip

\begin{center}

{\bf Table 4}

\bigskip

Table for Kronecker products of $D_4$ Irreps.

\bigskip

\bigskip

\begin{tabular}{|c||c|c|c|c|c|}
\hline
$\otimes $ & 1$_{1}$ & 1$_{2}$ & 1$_{3}$ & 1$_{4}$ & 2 \\ \hline\hline
1$_{1}$ & 1$_{1}$ & 1$_{2}$ & 1$_{3}$ & 1$_{4}$ & 2 \\ \hline
1$_{2}\ $ & 1$_{2}$ & 1$_{1}$ & 1$_{4}$ & 1$_{3}$ & 2 \\ \hline 
1$_{3}$ & 1$_{3}$ & 1$_{4}$ & 1$_{1}$ & 1$_{2}$ & 2 \\ \hline
1$_{4}$ & 1$_{4}$ & 1$_{3}$ & 1$_{2}$ & 1$_{1}$ & 2 \\ \hline
2 & 2 & 2 & 2 & 2 & $1_{1}+1_{2}+1_{3}+1_{4}$ \\ \hline
\end{tabular}

\end{center}

\bigskip
\bigskip

\noindent We see that Table 4 is contained in Table 3 and
that, from \cite{books}, $D_4 \not\subset Q_8$.

\bigskip

\noindent Thus the case $T \not\subset T^{'}$ is not unique yet
it is important to recall this relationship of $T$ to $T^{'}$
when building models of particle theory based on the binary
tetrahedral group.

\newpage

\noindent{\it Direct Products, Semidirect Products and Central Extensions}

\bigskip
\bigskip

\noindent A general group extension can be represented by a short exact
sequence 

$$1 \rightarrow H \rightarrow G \rightarrow K \rightarrow 1$$

\noindent which expresses the conditions that $H$ is a normal subgroup
of $G$ and $ G/H = K $.  We can distinguish three classes of
extensions from the rest: direct products, semidirect products,
and central extensions.  

\bigskip
\bigskip

\noindent Direct products are extensions of a 
trivial sort: $ H $ and $ K $ are both realized as subgroups of $ G $, and
the elements of each subgroup commute with those of the other; 
more concisely, $ G=HK, [H,K]=1 $. 

\bigskip
\bigskip

\noindent A semidirect product is one
in which the first property holds, but the multiplication in $H$
is twisted by an action of the elements of $K$: 

\bigskip

$$ g = h_1 k_1 h_2 k_2 = h_1 (k_1 h_2 k_1^{-1}) k_1 k_2 
= h_1 h_2^{k_1} k_1 k_2 $$

\bigskip

\noindent where the notation $h_2^{k_1}$ is defined by the parenthesized 
expression.  

\bigskip
\bigskip

\noindent A central extension is characterized by the property that $ H $ 
is a subgroup of the center of $ G,\ Z(G) $ (which consists of elements
commuting with all of $ G $).   The point 
is that an extension can be both central and a semidirect product only
if it is in fact a direct product: since $ H \subset Z(G) $, $ K \subset G $
implies $ [H,K] = 1 $.  Thus $ SU(2) $ is a central
extension of $ SO(3) $, and if $ K\subset SO(3) $, the corresponding extension
$K\, ' \subset SU(2) $  contains $ K $ only if $ K $ is abelian. 

\bigskip
\bigskip

\noindent Concerning our explicit examples,
$T$ is not a subgroup of $T^{'}$ because $SO(3)$
is not a subgroup of $SU(2)$; $T^{'}$ is a double
cover of $T$ because $SU(2)$ is a double cover
of $SO(3)$; $T^{'}$ is a central extension of $T$
because $SU(2)$ is a central extension of $SO(3)$;
$T$ is a central quotient of $T^{'}$ because
$SO(3)$ is a central quotient of $SU(2)$.
The same statements hold {\it mutatis mutandis}
replacing $T, T^{'}$ respectively by
$D_n, Q_{2n}$.

\newpage

\noindent {\it Discussion}

\bigskip
\bigskip

\noindent The present discussion can stand 
on its own as an original contribution to the
mathematics of group theory.

\bigskip

\noindent It is also germane to model building for
quarks and leptons.
As an example, consider the $T'$ model of quark masses
and neutrino masses and mixings discussed in \cite{FramptonKephart}.
There the leptons are in  singlets and triplets of $T'$, while the quarks
are in singlets and doublets. If for some reason one wished to break
the $T'$ quark mass
relations without disrupting the neutrino sector, i.e., without
disturbing tribimaximal mixing (TBM), one could hope to give a VEV to
a scalar in a nontrivial $T'$ irrep. The only possibility would appear
to be a $T'$ doublet
VEV. However, VEVs for doublets do not leave an unbroken $A_4$ symmetry to
protect TBM. Furthermore, not all $T'$ irreps can be decomposed into
$A_4$ irreps, as can easily be seen from studying the character tables of the
two groups. (As a specific example, the $2'$ is such a $T'$ irrep.)
This means that in the $T'$ model there is no consistent way to reduce the
quark symmetry, while temporarily preserving TBM. Other models can be
analysed in a similar fashion.

\bigskip

\noindent As evidenced by
\cite{catchall}, the groups $D_n, Q_{2n}, T, T^{'}, etc.$ have been used
extensively as flavor symmetries in the literature.
Many of these papers contain too hasty statements 
which can in the future be avoided so thereby
the present article contributes to
embedding nonabelian finite flavor
groups in continuous groups.

\bigskip
\bigskip
\bigskip
\bigskip
\bigskip
\bigskip
\bigskip
\bigskip

\begin{center}

{\bf Acknowledgements}

\end{center}

\bigskip

\noindent
This work was supported by U.S. Department of Energy grants number DE-FG02-06ER41418 
and  DE-FG05-85ER40226.

\end{document}